\documentstyle[12pt]{article}

\begin{document}

\setlength{\baselineskip}{18pt}

\begin{center}
{\large \bf Axial Monopoles, Quantization of Electric Charge and Dynamical
Discreteness of Space-Time\footnote{Work partially
supported by CNPq and FAPESP}}
\end{center}

\begin{center}
{\bf S.C.S. Silva$^{(1)}$, P.C.R.C. Mello$^{(2)}$ and M.C. Nemes$^{(2)}$}
\end{center}

\begin{center}
$^{(1)}$Instituto de F\'{\i}sica, Universidade de S\~ao Paulo\\
CP20516, 01498, S\~ao Paulo, SP, Brasil.
\end{center}

\begin{center}
$^{(2)}$Departamento de F\'{\i}sica, Universidade Federal de Minas
Gerais\\
CP702, 30000,
Belo Horizonte, MG, Brasil.
\end{center}

\begin{abstract}
In the present contribution we show that the introduction of
a conserved axial current
in electrodynamics can explain the quantization of electric charge,
preserving parity conservation, and
introduces a dynamical discreteness into space-time.
\end{abstract}

\noindent {\bf 1 - Introduction}

In 1931 Dirac proposes an electromagnetic theory
with magnetic monopoles$^{[1]}$, whose appeal is mainly connected
to  the possibility of explaining the quantization of the
electric charge. In spite of this undeniable
theoretical appeal, in Dirac's theory one is
faced with a symmetry problem: the terms responsible for the
magnetic monopole in
the generalized Maxwell's equations violate their symmetry under space
and time reversal.

In this work we propose the introduction of a new current, namely an {\it axial
electromagnetic current} which presents the following differences as compared
to previously proposed vector magnetic current:

\begin{description}

\item $a$) the resulting theory preserves space and time inversion
invariance;

\item $b$) besides the usual conservation of the vector electromagnetic
current, we also have the conservation of an axial current;

\item $c$) besides the charge quantization, we can obtain a dynamical
discreteness of space-time;

\end{description}

\noindent {\bf 2 - Axial photons and the axial electromagnetic current}

We start with the generalized definition of the electromagnetic field
tensor

\begin{equation}
F_{\mu\nu}=\partial_{\mu}A_{\nu}-\partial_{\nu}A_{\mu}
+\epsilon_{\mu\nu\alpha\beta}\partial^{\alpha}B^{\beta}
\end{equation}

\noindent where $B^{\mu}$ represents a new gauge field$^{[2]}$.
Maxwell's equations for the fields $A^{\mu}$ and $B^{\mu}$ in
Lorenz's gauge $(\partial^{\mu}A_{\mu}=\partial^{\mu}B_{\mu}=0)$
become

\begin{eqnarray}
\partial^{\nu}F_{\nu\mu}=\Box A_{\mu}=j_{\mu}\\
\partial^{\nu}F_{\nu\mu}^{\dagger}=\Box B_{\mu}=g_{\mu}
\end{eqnarray}

\noindent where $F_{\nu\mu}^{\dagger}$ corresponds to $F_{\nu\mu}$'s dual
tensor.

The quantity $F_{\mu\nu}$ in (1) is a tensor;
$\epsilon_{\mu\nu\alpha\beta}$ is a pseudo-tensor and therefore the
field $B_{\mu}$ must be a pseudo-vector or an axial field. From the
point of view of quantum theory the field $B_{\mu}$ represents
photon-like particles except for $P$, $T$ and $C$ parities. In other
words, {\it axial photons}. From
this it follows  that $(3)$ is not invariant under
time and space reversal, unless $g^{\mu}$ is also a pseudo-vector.
For this reason, we shall introduce the axial electromagnetic current
given by

\begin{equation}
g_{\mu}=-g\bar{\psi}\gamma_{\mu}\gamma_{5}\psi
\end{equation}

\noindent where $\psi$ represents a spin $1/2$ particle
({\it axial monopole}) with axial charge g.

Since $F_{\nu\mu}^{\dagger}$ is antisymmetric one gets from $(3)$

\begin{equation}
\partial^\mu g_\mu =0
\end{equation}

\noindent
which means axial current conservation and therefore massless
axial monopoles.

\noindent {\bf 3 - Charge quantization}

Let us analyze the compatibility of the axial electromagnetic current
with charge quantization.

We consider the gauge-invariant wave function of a charged particle$^{[2]}$
moving in the presence of the axial monopole's
electromagnetic field,

\begin{eqnarray}
\Phi_{e}\left(x,P'\right)  & = & \Phi_{e}\left(x,P\right)
\exp\left[-\frac{\imath e}{2}\int_{S} F^{\mu\nu}d\sigma_{\mu\nu}
\right]
\end{eqnarray}

\noindent $S$ being any surface with contour $P'-P$.
Due to the arbitrariness of the surface $S$ we can write

\begin{equation}
\Phi_{e}\left(x,P\right)
\exp\left[-\frac{\imath e}{2}\int_{S} F^{\mu\nu}d\sigma_{\mu\nu}
\right]=
\Phi_{e}\left(x,P\right)
\exp\left[-\frac{\imath e}{2}\int_{S'} F^{\mu\nu}d\sigma_{\mu\nu}
\right]
\end{equation}

\noindent which leads to the condition

\begin{equation}
\exp\left[-\frac{\imath e}{2}\oint_{S-S'} F^{\mu\nu}d\sigma_{\mu\nu}
\right]=1
\end{equation}

\noindent or equivalently to

\begin{equation}
\exp\left[-\imath e\int_{V} \partial^{\nu}F_{\nu\mu}^{\dagger}dV^{\mu}
\right]=1
\end{equation}

\noindent where $V$ is the volume involved by the closed
surface $S-S'$.

Using $(3)$, we have

\begin{equation}
\exp\left[-\imath e\int_{V}g_{\mu}dV^{\mu}\right]=1
\end{equation}

\noindent which gives

\begin{equation}
Q_V \equiv \int_V g_{\mu} dV^{\mu} = \frac{2 \pi n}{e}
\end{equation}

\noindent $n$ being any integer.

Using our definition of $g_{\mu}$ $(4)$, we get

\begin{equation}
Q_{V}=\int_{V}\left(-g\bar{\psi}\gamma_{\mu}\gamma_{5}\psi\right)
dV^{\mu}
\end{equation}

As $Q_{V}$ is a Lorentz scalar, we can perform the calculation in a
convenient reference frame. Taking the axial monopole's frame
(we can do that formally, even axial monopole being massless) and
using the standard representation for Dirac's spinor

\begin{equation}
\psi = \left(
\begin{array}{c}
\phi\\
\chi
\end{array}
\right)=
\left(
\begin{array}{c}
\phi\\
0
\end{array}
\right)
\end{equation}

\noindent we obtain

\begin{equation}
Q_V = g \int_V (\phi^{\dagger}\;\sigma_i\;\phi)\;dV^i
\end{equation}

\noindent where $\sigma_i$ corresponds to the $i^{th}$ Pauli matrix.

Taking now the axial
monopole polarization axis in the direction of charge's
velocity (positive z-axis, say) we get

\begin{equation}
Q_V = g \int \phi^{\dagger}\;\phi\;dx\;dy\;dt
\end{equation}

Since the axial monopole must be massless, the charge's velocity relative
to it has to be  necessarily $1$, the velocity of light. Thus $dt=dz$ and
$(15)$ leads to

\begin{equation}
Q_V = g \int
\phi^{\dagger}\;\phi\;dx\;dy\;dz= g
\end{equation}

Equations $(11)$ and $(16)$ give

\begin{equation}
\frac{eg}{2\pi} = n
\end{equation}

The main feature of this condition is that it does not depend on the
distance between the electric charge and the axial monopole.
It implies in charge quantization, in the same way of Dirac's
charge quantization condition$^{[1]}$.

\noindent {\bf 4 - Axial monopole's mass generation}

As we have shown (see $(5)$), in the present context the axial monopole
is necessarily massless, oppositely to the massive gauge solitonic
description of magnetic monopoles$^{[3]}$. Let us see what happens if
we circunvent such restriction.

We shall do that through a dynamical mass generation mechanism,
by introducing a Higgs scalar field with nonvanishing vacuum
expectation value$^{[4]}$. The Yukawa coupling between the Higgs
field and the axial
monopole will generate a mass term for the latter, but preserving the axial
current conservation $(5)$.

The free lagrangian for the massless axial monopole is

\begin{equation}
{\cal L}_0 = i \bar{\psi} \partial_{\mu} \gamma^{\mu} \psi
\end{equation}

\noindent
This lagrangian is invariant under the $U_A(1)$ transformations defined by

\begin{equation}
U_A(1):\; \psi \rightarrow e^{i\alpha\gamma_5} \psi
\end{equation}

\noindent
and this invariance leads to the axial current conservation $(5)$.

Now, we add to this free lagrangian the Higgs and Yukawa terms, to obtain

\begin{equation}
{\cal L} = {\cal L}_0 + {\cal L}_{Higgs} - G \bar{\psi}_L \phi_H \psi_R
- G \bar{\psi}_R \phi_H^{\dagger} \psi_L
\end{equation}

\noindent where $\phi_H$ stands for the Higgs scalar field, $\psi_L$ and
$\psi_R$ are, respectively, the left and right component of $\psi$ and
$G$ is the Yukawa coupling constant.

The lagrangian $(20)$ leads to a massive Dirac equation for the axial monopole
wave function $\psi$, with a mass term given by

\begin{equation}
M=GV
\end{equation}

\noindent with $V$ standing for the vacuum expectation value of the Higgs
field.

It is easy to see that ${\cal L}$ is invariant under $U_A(1)$ transformations,
since they transform the Higgs field as

\begin{equation}
\phi_H \rightarrow e^{-i\alpha} \phi_H
\end{equation}

Using Noether's theorem we can obtain the conserved current associated to
this invariance. It is precisely our axial current $(4)$.
\vspace{0.5cm}

\noindent {\bf 5 - Dynamical discreteness of space-time}

Let us investigate the consequences of the mass generation on the
electric charge quantization condition.

If the axial monopole is not massless, the charge's velocity relative to it,
$v$, is not necessarily $1$. Now, $dt=dz/v$ and from $(15)$ we obtain

\begin{equation}
Q_V = \frac{g}{v} \int
\phi^{\dagger}\;\phi\;dx\;dy\;dz= \frac{g}{v}
\end{equation}

Inserting $(23)$ in $(11)$ we have, rather than $(17)$, the condition

\begin{equation}
\frac{eg}{2\pi v} = n
\end{equation}

\noindent
that, again, is independent on the distance between the electric charge and
the axial monopole.

The above relation can be satisfied if we simultaneously fulfill

\begin{equation}
\frac{eg}{2\pi} = n_0
\end{equation}

\noindent and

\begin{equation}
v=\frac{n_0}{n}
\end{equation}

\noindent with

\begin{equation}
n\;=\;n_0,\;n_0+1,\;n_0+2...
\end{equation}

\noindent
$n_0$ being an integer fixed by the values of $e$ and $g$.

Equation $(25)$ is the charge quantization condition $(17)$,
already derived in the massless case. It can be formally obtained from
$(24)$ if we consider the limit in which the mass of the particle carrying
electric charge vanishes. Or, in another way, if we ``switch off" the axial
monopole mass, taking the false Higgs vacuum, in which $V=0$.
Physically we do not expect charge quantization to depend on the mass
of the particles or on any mass generation mechanism. We shall
therefore assume $(25)-(27)$ as the only physical solution of $(24)$.

Condition $(26)$ restricts the values of charge's velocity to rational
numbers, a result integrated in the theories of discrete space-time$^{[5]}$.
Besides, these rational values form a discrete sequence given by $(27)$.
For sufficiently high $n_0$, this sequence tends to a continuum, except
for velocities very near $1$, the light's velocity.

If we consider a massive charged particle, equations $(26)$ and $(27)$
lead to an upper limit for the particle's velocity, given by

\begin{equation}
v_0=\frac{n_0}{n_0+1}<1
\end{equation}

\noindent
since for a massive particle it is impossible to have $v=1$. For $n_0\gg1$
the limit  $v_0$ is very near $1$.

This limitation of $v$ leads to upper limits for $p$ and $E$, the momentum and
energy of such particle. For $n_0\gg1$ these upper limits are

\begin{equation}
p_0\approx E_0\approx m(n_0/2)^{\frac{1}{2}}
\end{equation}

\noindent
which are proportional to the particle's mass, $m$.

The limitation of the energy-momentum space of the particle leads, through
the uncertainty principle, to a discreteness of its space-time, with a
fundamental length given by

\begin{equation}
a \sim \frac{1}{p_0} \approx \frac{(2/n_0)^\frac{1}{2}}{m}
\end{equation}

The discreteness of space-time here has a dynamical nature, opposed
to usual theories of discrete space-time$^{[6]}$ where it is purely
kinematical in origin. Here a fundamental
length arises owing to the interaction between the charged particle and the
axial monopole. Besides, the larger the electric charge's mass the
smaller the fundamental lenght $a$,
such that in the classical limit ($m\gg0$) space-time will tend to
a continuum.
This kind of discreteness we shall call {\it dynamical discreteness}.

The experimental upper limit $a<10^{-16}cm$ for the fundamental length of
space-time gives a lower limit for $n_0$. Inserting such limit in $(30)$
and using for $m$ the electron's mass, we get

\begin{equation}
n_0>2\times10^6\gg1
\end{equation}

Now, from $(25)$ and using for $e$ the electron's charge, we obtain a
very high lower
limit for the axial charge $g$

\begin{equation}
g>10^9
\end{equation}

\noindent {\bf 6 - Conclusion}

The introduction of the concept of axial monopoles opens up
several interesting theoretical perspectives. Well known symmetries
in nature, such as space and time reversal, are preserved and a new
symmetry, namely the $U_A(1)$ symmetry, arises in the context of
electrodynamics. Charge quantization can be obtained and, in the
massive case, it is shown to be intimately connected to a dynamical
discreteness of space-time.

Now an important remark is in order: what can we say about the
observation of such entity? Firstly we can say nothing definite
about its mass, since the values for $G$ and $V$ in $(21)$ are
unknown. However, what is known is that producing massive pairs with
large opposite coupling constants (see $(32)$) may be a
very difficult experimental task. Moreover, the coupling of the axial
monopole to the electromagnetic field is not of vector character, but
axial (see $(3)$ and $(4)$). This means in particular that its
production and detection
will be directly associated with polarization conditions, which will
render its observation non trivial, even in the massless case.

\end{document}